\documentclass{Interspeech2024}

\interspeechcameraready

\usepackage{mathtools}
\usepackage{subcaption}
\usepackage{hyperref}
\usepackage{cleveref}
\usepackage{amsmath}
\usepackage{xcolor}
\usepackage{hyperref}
\usepackage{color}
\newcommand{\loss}[1]{\mathcal{L}_{#1}}
\newcommand{\lam}[1]{\lambda_{#1}}

\let\svthefootnote\thefootnote
\newcommand\freefootnote[1]{%
  \let\thefootnote\relax%
  \footnotetext{#1}%
  \let\thefootnote\svthefootnote%
}

\title{FreeV: Free Lunch For Vocoders Through Pseudo Inversed Mel Filter}

\name[affiliation={1}]{Yuanjun}{Lv}
\name[affiliation={2}]{Hai}{Li}
\name[affiliation={2}]{Ying}{Yan}
\name[affiliation={2}]{Junhui}{Liu}
\name[affiliation={2}]{Danming}{Xie}
\name[affiliation={1,*}]{Lei}{Xie}

\address{
  $^1$Audio, Speech and Language Processing Group (ASLP@NPU), School of Computer Science, \\Northwestern Polytechnical University, Xi’an, China\\
  $^2$iQIYI Inc., Chengdu, China
}
\email{yjlv@mail.nwpu.edu.cn,\
lxie@nwpu.edu.cn}


\begin{document}

\maketitle

\freefootnote{*: corresponding author}
\begin{abstract}
    

Vocoders reconstruct speech waveforms from acoustic features and play a pivotal role in modern TTS systems. Frequent-domain GAN vocoders like Vocos and APNet2 have recently seen rapid advancements, outperforming time-domain models in inference speed while achieving comparable audio quality. However, these frequency-domain vocoders suffer from large parameter sizes, thus introducing extra memory burden. Inspired by PriorGrad and SpecGrad, we employ pseudo-inverse to estimate the amplitude spectrum as the initialization roughly. This simple initialization significantly mitigates the parameter demand for vocoder. Based on APNet2 and our streamlined Amplitude prediction branch, we propose our FreeV, compared with its counterpart APNet2, our FreeV achieves \textbf{1.8$\times$ inference speed improvement} with nearly \textbf{half parameters}. Meanwhile, our FreeV outperforms APNet2 in resynthesis quality, marking a step forward in pursuing real-time, high-fidelity speech synthesis. Code and checkpoints is available at: \url{https://github.com/BakerBunker/FreeV}

\end{abstract}
\vspace{-0.1in}

\section{Introduction}

Recently, there has been a rapid advancement in the field of neural vocoders, which transform speech acoustic features into waveforms. These vocoders play a crucial role in text-to-speech synthesis, voice conversion, and audio enhancement applications. Within these contexts, the process typically involves a model that predicts a mel-spectrogram from the source text or speech, followed by a vocoder that produces the waveform from the predicted mel-spectrogram. Consequently, the quality of the synthesized speech, the speed of inference, and the parameter size of the model constitute the three primary metrics for assessing the performance of neural vocoders.

Recent advancements in vocoders, including iSTFTNet~\cite{kaneko2022istftnet}, Vocos~\cite{siuzdak2023vocos}, and APNet~\cite{ai2023apnet}, have shifted from the prediction of waveforms in the time domain to the estimation of amplitude and phase spectra in the frequency domain, followed by waveform reconstruction via inverse short-time Fourier transform (ISTFT). This method circumvents the need to predict extensive time-domain waveforms, thus reducing the models' computational burden. ISTFTNet, for example, minimizes the computational complexity by decreasing the upsampling stages and focusing on frequency-domain spectra predictions before employing ISTFT for time-domain signal reconstruction. Vocos extends these advancements by removing all upsampling layers and utilizing the ConvNeXtV2~\cite{woo2023convnext} Block as its foundational layer. APNet~\cite{ai2023apnet} and APNet2~\cite{du2023apnet} further refine this approach by independently predicting amplitude and phase spectra and incorporating innovative supervision to guide phase spectra estimation. Nonetheless, with comparable parameter counts, these models often underperform their time-domain counterparts, highlighting potential avenues for optimization in the parameter efficiency of frequency-domain vocoders.

Several diffusion-based vocoders have integrated signal-processing insights to reduce inference steps and improve reconstruction quality. PriorGrad~\cite{lee2021priorgrad} initially refines the model's priors by aligning the covariance matrix diagonals with the energy of each frame of the Mel spectrogram. Extending this innovation, SpecGrad~\cite{koizumi2022specgrad} proposed to adjust the diffusion noise to align its dynamic spectral characteristics with those of the conditioning mel spectrogram. Moreover, GLA-Grad~\cite{liu2024glagrad} enhances the perceived audio quality by embedding the estimated amplitude spectrum into each diffusion step's post-processing stage. Nevertheless, the reliance on diffusion models results in slower inference speeds, posing challenges for their real-world application.

\begin{figure}[t]
    \centering
    \includegraphics[width=0.8\linewidth]{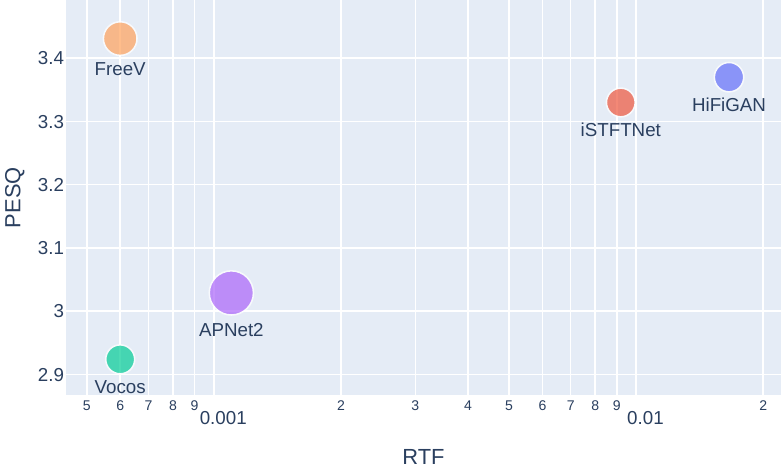}
    \vspace{-0.1in}
    \caption{Inference speed and reconstruction performance of multiple methods trained and evaluated on LJSpeech. The size of the circle represents the model parameter size. FreeV achieves the fastest inference speed and reconstruction quality with half parameter size compared to APNet2.}
    \label{fig:perf_compare}
    \vspace{-0.3in}
\end{figure}

In this work, we introduce \textit{FreeV}, a streamlined GAN vocoder enhanced with prior knowledge from signal processing, and tested on the LJSpeech dataset~\cite{ljspeech17}. The empirical outcomes highlight FreeV's superior performance characterized by faster convergence in training, a near 50\% reduction in parameter size, and a notable boost in inference speed. Our contributions can be summarized as follows:
\begin{itemize}
    \item We innovated by using the product of the Mel spectrogram and the pseudo-inverse of the Mel filter, referred to as the pseudo-amplitude spectrum, as the model's input, effectively easing the model's complexity.
    \item Drawing on our initial insight, we substantially diminished the spectral prediction branch's parameters and the time required for inference without compromising the quality achieved by the original model.
\end{itemize}

\section{Related Work}

\subsection{PriorGrad \& SpecGrad}
\label{sec:specgrad}
Based on diffusion-based vocoder WaveGrad~\cite{chen2021wavegrad}, which direct reconstruct the waveform through a DDPM process, Lee \textit{et al.} proposed PriorGrad~\cite{lee2021priorgrad} by introducing an adaptive prior $\mathcal{N}(\mathbf{0},\mathbf{\Sigma})$, where $\mathbf{\Sigma}$ is computed from input mel spectrogram $X$. The covariance matrix $\mathbf{\Sigma}$ is given by: $\mathbf{\Sigma}=\mathrm{diag} [(\sigma_1^2,\sigma_2^2,\cdots,\sigma_D^2,)]$, where $\sigma_d^2,$ denotes the signal power at $d$th sample, which is calculated by interpolating the frame energy.  Compared to conventional DDPM-based vocoders, PriorGrad utilizes signal before making the source distribution closer to the target distribution, which simplifies the reconstruction task.

Based on PriorGrad, SpecGrad~\cite{koizumi2022specgrad} proposed adjusting the diffusion noise in a way that aligns its dynamic spectral characteristics with those of the conditioning mel spectrogram. SpecGrad introduced a decomposed covariance matrix and its approximate inverse using the idea from T-F domain filtering, which is conditioned on the mel spectrogram. This method enhances audio fidelity, especially in high-frequency regions. We denote the STFT by a matrix $G$, and the ISTFT by a matrix $G^+$, then the time-varying filter $L$ can be expressed as:
\begin{equation}
    L=G^+DG,
\end{equation}
where $D$ is a diagonal matrix that defines the filter, and it is obtained from the spectral envelope. Then we can obtain covariance matrix $\Sigma=LL^T$ of the standard Gaussian noise $\mathcal{N}(0,\Sigma)$ in the diffusion process. By introducing more accurate prior to the model, SpecGrad achieves higher reconstruction quality and inference speech than PriorGrad.
\subsection{APNet \& APNet2}
\label{sec:apnet}
\begin{figure}[t]
    \centering
    \includegraphics[width=0.7\linewidth]{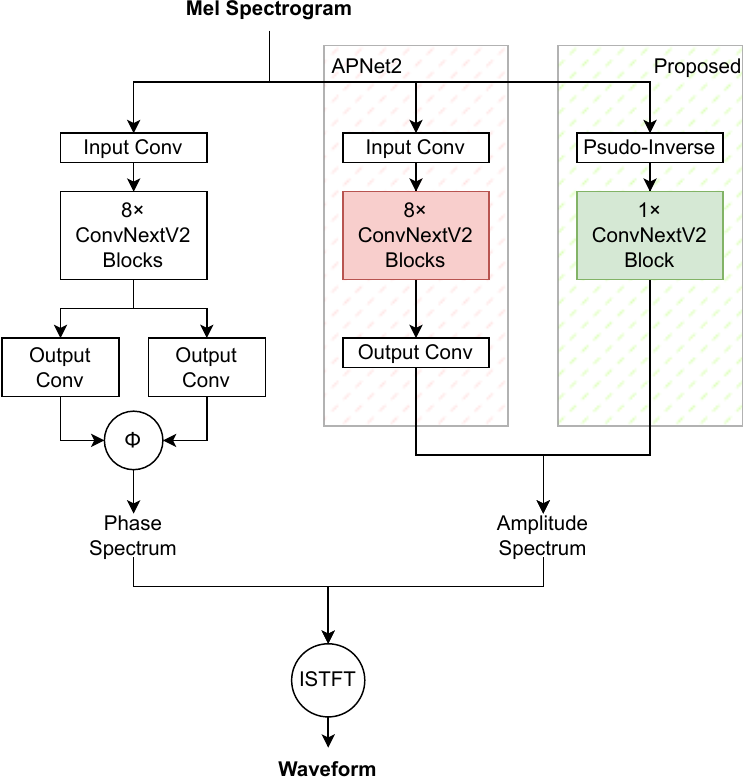}
    \caption{The overall architecture of FreeV, the amplitude prediction branch (ASP) of APNet2, which has \textcolor{red}{red} background, is replaced by a more lightweight architecture with \textcolor{green}{green} background.}
    \label{fig:overall}
    \vspace{-0.2in}
\end{figure}

As illustrated in Figure~\ref{fig:overall}, APNet2~\cite{du2023apnet} consists of two components: amplitude spectra predictor (ASP) and phase spectra predictor (PSP). These two components predict the amplitude and phase spectra separately, which are then employed to reconstruct the waveform through ISTFT. The backbone of APNet2 is ConvNeXtV2~\cite{woo2023convnext} block, which is proved has strong modeling capability. In the PSP branch, APNet~\cite{ai2023apnet} proposed the parallel phase estimation architecture at the output end. The parallel phase estimation takes the output of two convolution layers as the pseudo imaginary part $I$ and real part $R$, then obtains the phase spectra by:
\begin{equation}
    \arctan(\frac{I}{R})-\frac{\pi}{2}\cdot sgn(I)\cdot[sgn(R)-1]
\end{equation}
where $sgn$ is the sign function.

A series of losses are defined in APNet to supervise the generated spectra and waveform. In addition to the losses used in HiFiGAN~\cite{kong2020hifigan}, which include Mel loss $\loss{mel}$, generator loss $\loss{g}$, discriminator loss $\loss{d}$, feature matching loss $\loss{fm}$, APNet proposed: 
\begin{itemize}
    \item amplitude spectrum loss $\loss{A}$, which is the L2 distance of the predicted and real amplitude;
    \item  phase spectrogram loss $\loss{P}$, which is the sum of instantaneous phase loss, group delay loss, and phase time difference loss, all phase spectrograms are anti-wrapped;
    \item STFT spectrogram loss $\loss{S}$, which includes the STFT consistency loss and L1 loss between predicted and real reconstructed STFT spectrogram.
\end{itemize}
\section{Method}
\label{sec:method}
\begin{flushright}
\textit{``When we structure the informative prior noise closer to the data distribution, can we improve the efﬁciency of the model?"  -- PriorGrad}
\vspace{-0.2in}
\end{flushright}

\subsection{Amplitude Prior}
\label{sec:prior}
In this section, we investigate how to obtain a prior signal closer to the real prediction target, which is the amplitude spectrum. By employing the given Mel spectrum $X$ and the known Mel filter $M$, we aim to obtain an amplitude spectrum $\hat{A}$ that minimizes the distance with the actual amplitude spectrum $A$, while ensuring that the computation is performed with optimal speed, as the following equation:
\begin{equation}
    \min\left\lVert \hat{A}M-A \right\rVert_2
\end{equation}
We investigated several existing implementations for this task. In Section \ref{sec:specgrad}, the SpecGrad method, $G^+DG\epsilon$ requires prior noise $\epsilon$ as input, therefore unsuitable for our goals. In the implementation by the librosa library~\cite{librosa}, the estimation of $\hat{A}$ employs the Non-Negative Least Squares (NNLS) algorithm to maintain non-negativity. However, this algorithm is slow due to the need for multiple iterations, prompting the pursuit of a swifter alternative. TorchAudio's implementation~\cite{torchaudio} calculates the estimated amplitude spectrum through a singular least squares operation followed by enforcing a minimum threshold to preserve non-negativity. Despite this, the recurring need for the least squares calculation with each inference introduces speed inefficiencies.

Considering that the Mel filter $M$ remains unchanged throughout the calculations, we can pre-compute its pseudo-inverse, denoted as $M^+$. Then, to guarantee the non-negativity of the amplitude spectrum and maintain numerical stability in training, we impose a lower bound of $10^{-5}$ on the values of the approximate amplitude spectrum.
\begin{figure}
    \centering
    \centering
    \begin{subfigure}[b]{0.3\columnwidth}
        \centering
        \includegraphics[width=\linewidth]{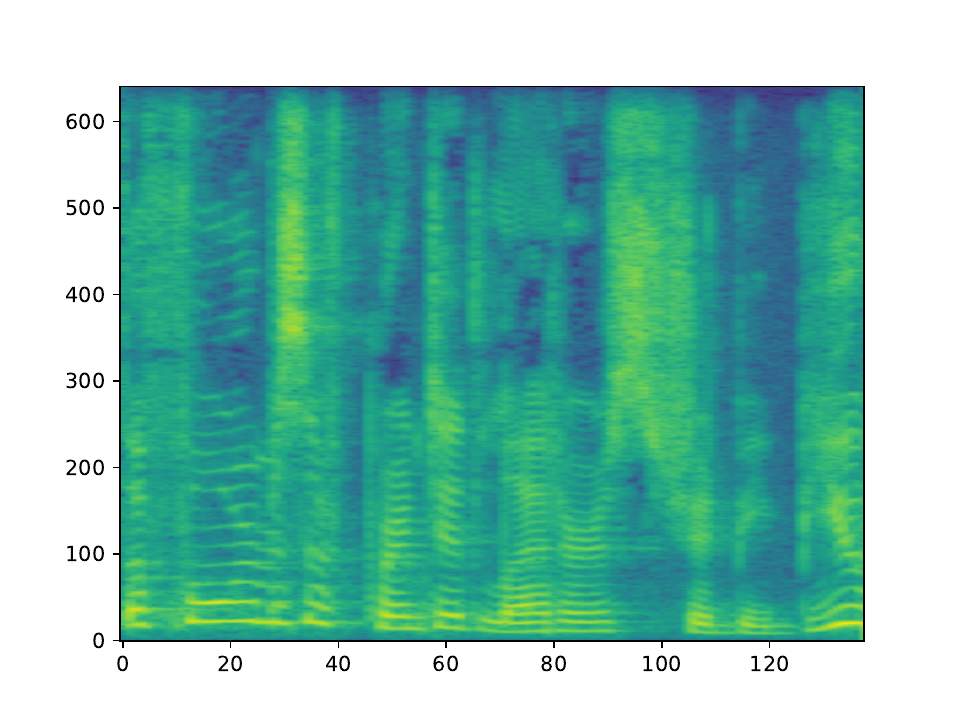}
        \caption{$A$}
    \end{subfigure}
    \hfill
    \begin{subfigure}[b]{0.3\columnwidth}
        \centering
        \includegraphics[width=\linewidth]{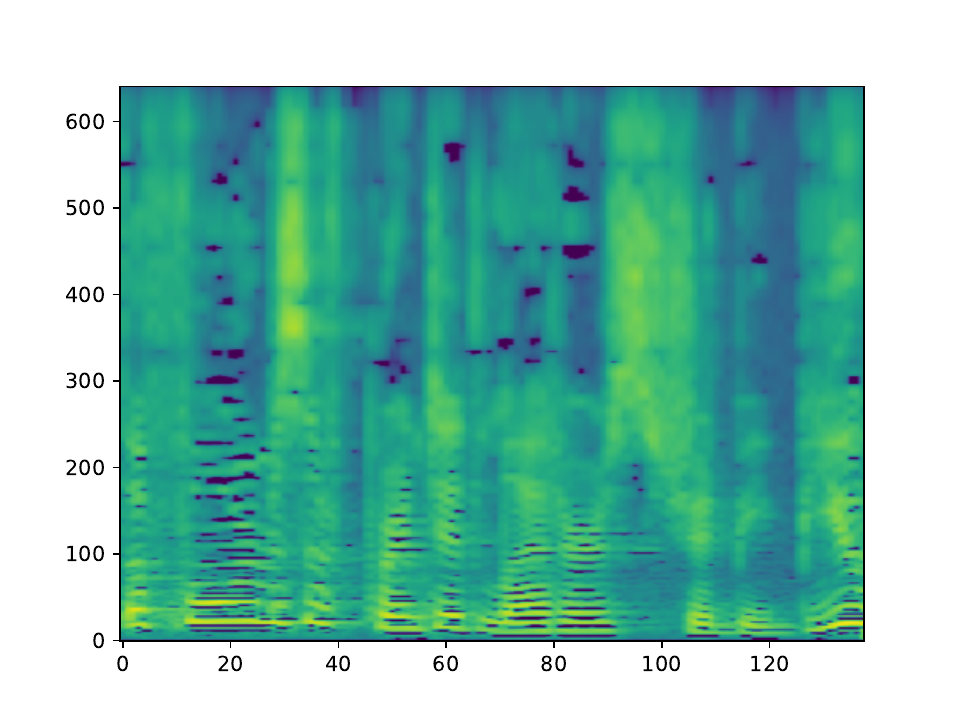}
        \caption{$\hat{A}$ w/o abs}
        \label{subfig:recon_wo_abs}
    \end{subfigure}
    \hfill
    \begin{subfigure}[b]{0.3\columnwidth}
        \centering
        \includegraphics[width=\linewidth]{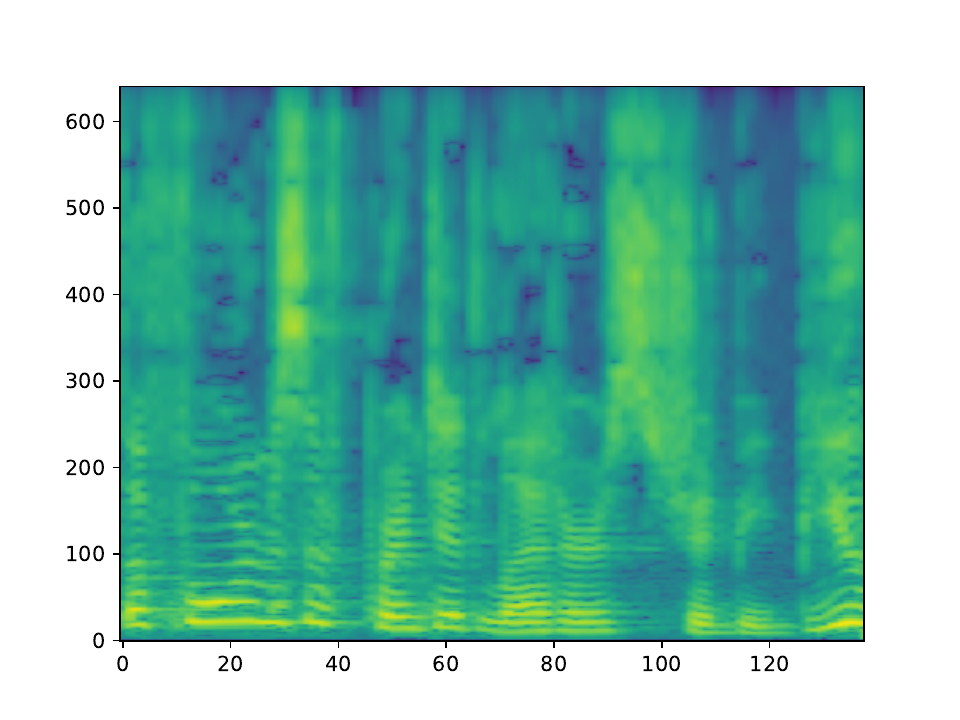}
        \caption{$\hat{A}$ w/ abs}
    \end{subfigure}
    \caption{Comparison of real log amplitude spectra $A$ and estimated log spectra $\hat{A}$.}
    \vspace{-0.2in}
    \label{fig:amp_compare}
\end{figure}
We find there are some negative values in the pseudo-inversed mel filter, causing negative blocks in estimated amplitude, which can be easily found in Figure \ref{subfig:recon_wo_abs}, so we add an $\mathrm{Abs}$ function to the product of $M^+$ and $X$.
This allows us to derive the approximate amplitude spectrum $\hat{A}$ using the following equation:
\begin{equation}
    \hat{A}=\mathrm{max}(\mathrm{Abs}(M^+X),10^{-5})
\end{equation}
 This enables us to efficiently acquire the estimated amplitude spectrum through a single matrix multiplication operation.
\subsection{Model Structure}
Our model architecture is illustrated in Figure 2, which consists of PSP and ASP, and uses ConvNextV2~\cite{woo2023convnext} as the model's basic block. PSP includes an input convolutional layer, eight ConvNeXtV2 blocks, and two convolutional layers for parallel phase estimation structure. 

Diverging from APNet2's ASP, our design substitutes the conventional input convolutional layer with the pre-computed pseudo-inverse Mel filter matrix $M^+$ of the Mel filter $M$ with frozen parameters. Due to the enhancements highlighted in Section \ref{sec:prior} that substantially ease the model's complexity, the number of ConvNeXtV2 blocks is reduced from eight to a single block, thereby substantially reducing both the parameter footprint and computation time. 

Concurrently, the ConvNeXtV2 module's input-output dimensions have been tailored to align with those of the amplitude spectrum, enabling the block to exclusively model the residual between the estimated and real amplitude spectra, further reducing the ASP module's modeling difficulty. Because the input and output dimensions of the ConvNeXtV2 module match the amplitude spectrum, we removed the output convolutional layer from ASP, further reducing the model's parameter count.

\subsection{Training Criteria}
In the choice of discriminators, we followed the setup in APNet2~\cite{du2023apnet}, using MPD and MRD as discriminators and adopting Hinge GAN Loss as the loss function for adversarial learning. We also retained the other loss functions used by APNet2, which is described in Section \ref{sec:apnet}, and the loss function of the generator and discriminator are denoted as:
\begin{align*}
    \vspace{-0.2in}
    \loss{Gen}&=\lam{A}\loss{A}+\lam{P}\loss{P}+\lam{S}\loss{S}+\lam{W}(\loss{mel}+\loss{fm}+\loss{g}) \\
    \loss{Dis}&=\loss{d}
    \vspace{-0.2in}
\end{align*}
where $\lam{A}$, $\lam{P}$, $\lam{S}$, $\lam{W}$ are the weights of the loss, which are kept the same as in APNet2.

\section{Experimental Setup}
To evaluate the effectiveness of our proposed FreeV, we follow the training scheme in APNet2 paper. Our demos are placed at demo-site\footnote{\url{https://bakerbunker.github.io/FreeV/}}.
\subsection{Dataset}
\label{sec:dataset}
To ensure consistency, the training dataset follows the same configuration of APNet2. Thus, the LJSpeech dataset~\cite{ljspeech17} is used for training and evaluation. LJSpeech dataset is a public collection of 13,100 short audio clips featuring a single speaker reading passages from 7 non-fiction books. The duration of the clips ranges from 1 to 10 seconds, resulting in a total length of approximately 24 hours. The sampling rate is 22050Hz. We split the dataset to train, validation, and test sets according to open-source VITS repository\footnote{\url{https://github.com/jaywalnut310/vits/tree/main/filelists}}.

For feature extraction, we use STFT with 1024 bins, a hop size of 256, and a Hann window of length 1024. For the mel filterbank, 80 filterbanks are used with a higher frequency cutoff at 16 kHz. 
\subsection{Model and Training Setup}
\label{sec:settings}
We compare our proposed model with HiFiGAN\footnote{\url{https://github.com/jik876/hifi-gan}}~\cite{kong2020hifigan}, iSTFTNet\footnote{\url{https://github.com/rishikksh20/iSTFTNet-pytorch}}~\cite{kaneko2022istftnet}, Vocos\footnote{\url{https://github.com/gemelo-ai/vocos}}~\cite{siuzdak2023vocos}\ and APNet2\footnote{\url{https://github.com/redmist328/APNet2}}~\cite{du2023apnet}. In Our FreeV vocoder, the number of ConvNeXtV2 blocks is 8 for PSP and 1 for ASP, the input-output dimension is 512 for PSP and 513 for ASP, the hidden dimension is 1536 for both ASP and PSP.

We trained FreeV for 1 million steps. We set the segmentation size to 8192 and the batch size to 16. We use the AdamW optimizer with $\beta_1=0.8$, $\beta_2=0.99$, and a weight decay of 0.01. The learning rate is set to $2\times10^{-4}$ and exponentially decays with a factor of 0.99 for each epoch.

\subsection{Evaluation}
Multiple objective evaluations are conducted to compare the performance of these vocoders. We use seven objective metrics for evaluating the quality of reconstructed speech, including mel-cepstrum distortion (MCD), root mean square error of log amplitude spectra and F0 (LAS-RMSE and F0-RMSE), V/UV F1 for voice and unvoiced part, short time objective intelligence (STOI)~\cite{stoi} and perceptual evaluation speech quality (PESQ)~\cite{pesq}. To evaluate the efficiency of each vocoder, model parameter count (Params) and real-time factor (RTF) are also conducted on NVIDIA A100 for GPU and a single core of Intel Xeon Platinum 8369B for CPU.

For the computational efficiency of the prior, we also conducted RTF and LAS-RMSE evaluations to the NNLS algorithm of librosa, least square algorithm of torchaudio, pseudo-inverse algorithm, and pseudo-inverse algorithm with absolute function mentioned in Section \ref{sec:prior}.
\begin{table}[ht]
\caption{Time and precision of different prior computing methods, LS stands for Least Square, PI stands for Pseudo Inverse.}
\label{tab:prior_compute}
\begin{tabular}{l|llll}
\toprule
Method       & NNLS   & LS        & PI        & PI w/ abs \\ \midrule
Time (↓)     & 290ms  & 286$\mu$s & 102$\mu$s & 107$\mu$s \\
LAS-RMSE (↓) & 2.0729 & 2.0729    & 2.0729    & 0.6843    \\ \bottomrule
\end{tabular}
\vspace{-0.2in}
\end{table}
\begin{table*}[ht]
\caption{Results of objective evaluations on the testset of LJSpeech dataset for reconstruction.}
\label{tab:perf}
\begin{tabular}{l|ccccccc}
\toprule
Model                & MCD(↓)         & LAS-RMSE(↓)    & V/UV F1(↑)     & Periodicity (↓) & F0-RMSE(↓)     & STOI(↑)        & PESQ(↑)        \\ \midrule
HiFiGAN~\cite{kong2020hifigan}              & 3.857          & 1.150          & 0.941          & 0.145           & 36.03          & 0.923          & 3.370          \\
HiFiGAN w/$\hat{A}$  & 3.751          & 1.141          & 0.945          & 0.136           & 34.26          & 0.928          & \textbf{3.509} \\
iSTFTNet~\cite{kaneko2022istftnet}             & 3.838          & 1.130          & 0.941          & 0.143           & 36.40          & 0.925          & 3.330          \\
iSTFTNet w/$\hat{A}$ & 3.755          & 1.120          & 0.944          & 0.138           & 34.12          & 0.927          & 3.422          \\
Vocos~\cite{siuzdak2023vocos}                & 3.367          & 0.948          & 0.941          & 0.158           & 45.41          & 0.948          & 2.924          \\
APNet2~\cite{du2023apnet}               & 3.518          & 0.782          & 0.950          & 0.132           & 31.08          & 0.950          & 3.029          \\
FreeV (Proposed)     & \textbf{3.112} & \textbf{0.779} & \textbf{0.956} & \textbf{0.118}  & \textbf{26.40} & \textbf{0.967} & 3.431          \\ \bottomrule
\end{tabular}
\end{table*}
\begin{table}[h]
\caption{Results of parameter and inference speed.}
\label{tab:efficiency}
\begin{tabular}{l|ccc}
\toprule

Model            & Params (M)     & \begin{tabular}[c]{@{}c@{}}RTF\\ (CPU)\end{tabular} & \begin{tabular}[c]{@{}c@{}}RTF\\ (GPU)\end{tabular} \\ \midrule
HiFiGAN          & 13.9M          & 0.062                                               & 0.0166                                              \\
iSTFTNet         & \textbf{13.3M} & 0.409                                               & 0.0092                                              \\
Vocos            & 13.5M          & \textbf{0.028}                                      & \textbf{0.0006}                                     \\
APNet2           & 31.4M          & 0.062                                               & 0.0011                                              \\
FreeV (Proposed) & 18.2M          & 0.036                                               & \textbf{0.0006}  \\
\bottomrule
\end{tabular}
\end{table}
\section{Experiment Result}
We conducted experiments to verify whether our method can improve the efficiency of the vocoder.

\subsection{Computational Efficiency of Prior}
The compute method of the estimated amplitude spectra $\hat{A}$ if our key component. We find that the inference speed can be affected by the compute speed of the prior.
We compare the compute speed and accuracy on 100 2-second-long speech clips.
As shown in Table \ref{tab:prior_compute}, the pseudo-inverse method is the fastest way to compute the estimated amplitude spectra $\hat{A}$, and the result also shows that the $\mathrm{Abs}$ function can largely reduce the error of amplitude spectrogram estimation.

\subsection{Model Convergence}
\begin{figure}[h]
    \centering
    \begin{subfigure}[b]{0.3\columnwidth}
        \centering
        \includegraphics[width=\linewidth]{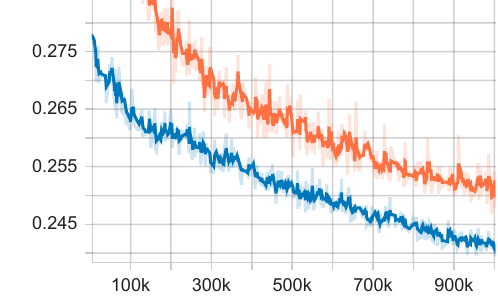}
        \caption{$\loss{mel}$}
        \label{subfig:AH_mel}
    \end{subfigure}
    \hfill
    \begin{subfigure}[b]{0.3\columnwidth}
        \centering
        \includegraphics[width=\linewidth]{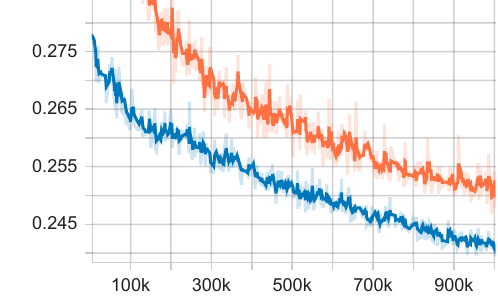}
        \caption{$\loss{amp}$}
        \label{subfig:AH_amp}
    \end{subfigure}
    \hfill
    \begin{subfigure}[b]{0.3\columnwidth}
        \centering
        \includegraphics[width=\linewidth]{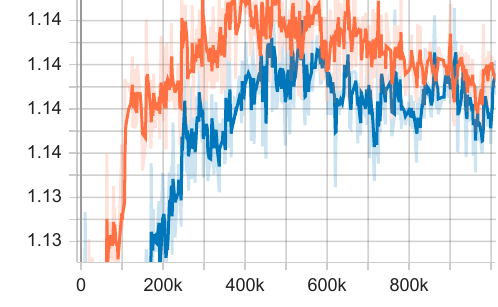}
        \caption{$\loss{ptd}$}
        \label{subfug:AH_ptd}
    \end{subfigure}
    \caption{Loss curves of APNet2 (\textcolor{orange}{orange}) and FreeV (\textcolor{blue}{blue}).}
    \label{fig:loss}
\end{figure}

In Figure \ref{subfig:AH_mel} and \ref{subfig:AH_amp}, we showcase the amplitude spectrum loss and mel spectrum loss curves related to amplitude spectrum prediction. From these two curves, it can be seen that even though the number of parameters in the amplitude spectrum prediction branch is significantly reduced, the loss related to amplitude spectrum prediction still remains lower than the baseline APNet2. This observation affirms the efficacy of the approach described in Section \ref{sec:method}, substantiating a marked decrease in the challenge of amplitude spectrum prediction. Furthermore, Figure \ref{subfug:AH_ptd} displays the Phase-Time Difference Loss, which bears significant relevance to phase spectrum prediction. The improvement in amplitude spectrum prediction concurrently benefits phase spectrum accuracy. We assume that the stability of the amplitude spectrum prediction branch's training engenders more effective optimization of the phase information by the waveform-related loss functions.

\begin{figure}[h]
    \begin{subfigure}[b]{0.3\columnwidth}
        \centering
        \includegraphics[width=\linewidth]{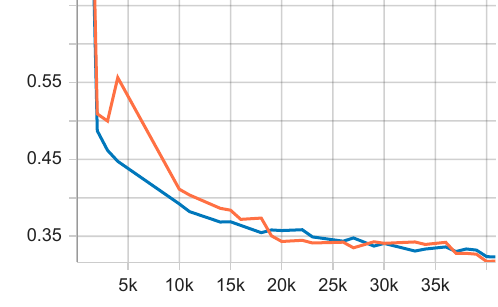}
        \caption{HiFiGAN}
    \end{subfigure}
    \hfill
    \begin{subfigure}[b]{0.3\columnwidth}
        \centering
        \includegraphics[width=\linewidth]{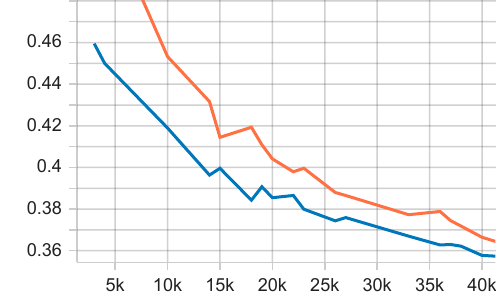}
        \caption{iSTFTNet}
    \end{subfigure}
    \hfill
    \begin{subfigure}[b]{0.3\columnwidth}
        \centering
        \includegraphics[width=\linewidth]{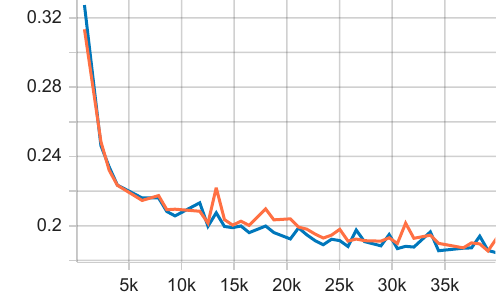}
        \caption{Vocos}
    \end{subfigure}
    \caption{Early stage mel loss curves of multiple models trained with (\textcolor{blue}{blue}) and without estimated amplitude spectra $\hat{A}$ (\textcolor{orange}{orange}).}
    \label{fig:loss_other}
    \vspace{-0.2in}
\end{figure}

Furthermore, we extended our experimentation to the baseline model by substituting its input from the Mel spectrum with the estimated amplitude spectrum $\hat{A}$. The loss curve illustrated in Figure \ref{fig:loss_other} reveals that this modification also enhanced the early-stage convergence of these models. This finding suggests that integrating an appropriate prior is advantageous not only for our proposed vocoder but also holds potential efficacy for other vocoder frameworks.

\subsection{Model Performance}
The model's performance was evaluated on the test dataset referenced in Section \ref{sec:dataset}, the results of which are detailed in Table \ref{tab:perf}. FreeV outperformed in five out of six objective metrics and was surpassed only by HiFiGAN with estimated amplitude spectra in the PESQ metric. These findings indicate that our method reduces the model's parameter size and elevates the quality of audio reconstruction. Furthermore, the comparative analysis, which includes both scenarios, with and without the incorporation of the estimated amplitude spectrum $\hat{A}$, reveals that substituting the Mel spectrum $X$ input with the approximate amplitude spectrum $\hat{A}$ can also yield performance gains in standard vocoder configurations. This observation corroborates the efficacy of our proposed approach.

In parallel, as shown by Table \ref{tab:efficiency}, our model's parameter size is confined to merely a half of that to APNet2, while it achieves 1.8$\times$ inference speed on GPU. When benchmarked against the time-domain prediction model HiFiGAN~\cite{kong2020hifigan}, FreeV not only exhibits a considerable speed enhancement, which is approximately 30$\times$, but also delivers superior audio reconstruction fidelity with comparable parameter count. These results further underscore the practicality and advantage of our proposed method.

\section{Conclusion}
In this paper, we investigated the effectiveness of employing pseudo-inverse to roughly estimate the amplitude spectrum as the initial input of the model. We introduce FreeV, a vocoder framework that leverages estimated amplitude spectrum $\hat{A}$ to simplify the model's predictive complexity. This approach not only reduces the parameter size but also improves the reconstruction quality compared to APNet2. Our experimental results demonstrated that our method could effectively reduce the modeling difficulty by simply replacing the input mel spectrogram with the estimated amplitude spectrum $\hat{A}$.

\newpage

\bibliographystyle{IEEEtran}
\bibliography{mybib}

\end{document}